\documentclass[12pt]{iopart}
\usepackage[pdftex]{graphicx}
\usepackage{iopams}
\usepackage{color}

\newcommand{\spin}[1]{\sigma_{#1}} % symbol for spins
\newcommand{\nn}[2]{\left<#1,#2\right>} % nearest neighbour spins
\newcommand{\nnn}[2]{\left<\left<#1,#2\right>\right>} 
\newcommand{\plaq}[4]{\left[#1,#2,#3,#4\right]} % plaquettes
\newcommand{\nnsum}{\sum\limits_{\nn{i}{j}}\spin{i}\spin{j}} 
\newcommand{\nnnsum}{\sum\limits_{\nnn{i}{j}}\spin{i}\spin{j}} 
\newcommand{\plaqsum}{\sum\limits_{\plaq{i}{j}{k}{l}}%
  \spin{i}\spin{j}\spin{k}\spin{l}}

\def\be{\begin{equation}}
\def\ee{\end{equation}}
\def\bea{\begin{eqnarray}}
\def\eea{\end{eqnarray}}

\begin{document}

%\markboth{D.~A.~Johnston, W.~Janke and M.~Mueller}{Macroscopic Degeneracy and Order in the $3d$ Plaquette Ising Model}

%\catchline{}{}{}{}{}
%
%%%%%%%%%%%%%%%%%%%%%%%%%%%%%%%%%%%%%%%%%%%%%%%%%%%%%%%%%%%%%%%%%%%%

\title{Macroscopic Degeneracy and order in the $3d$ plaquette Ising model}

\date{\today}

\author{Desmond A Johnston}

\address{School of Mathematical and Computer Sciences,
Heriot-Watt University,\\ Riccarton, Edinburgh EH14 4AS, Scotland.\\
D.A.Johnston@hw.ac.uk}

\author{Marco Mueller}

\address{Institut f\"ur Theoretische Physik, Universit\"at Leipzig,\\
   Postfach 100\,920, 04009 Leipzig, Germany.\\
    Marco.Mueller@itp.uni-leipzig.de}

\author{Wolfhard Janke}

\address{Institut f\"ur Theoretische Physik, Universit\"at Leipzig,\\
   Postfach 100\,920, 04009 Leipzig, Germany.\\
   Wolfhard.Janke@itp.uni-leipzig.de}

\begin{abstract}
The purely plaquette  $3d$ Ising Hamiltonian with the spins living at the
vertices of a cubic lattice  displays several interesting features. The
symmetries of the model lead to a macroscopic degeneracy of the low-temperature
phase and  prevent the definition of a standard magnetic order parameter.
Consideration of the strongly anisotropic limit of the model suggests that a
layered, ``fuki-nuke'' order still  exists and we confirm this with
multicanonical simulations. The macroscopic degeneracy of the low-temperature
phase also changes the finite-size scaling corrections at the first-order
transition in the model and we see this must be taken into account when
analysing our measurements.
\end{abstract}

\maketitle

%\keywords{Statistical mechanics; First-order phase transitions; Ising model.}

\section{Introduction}
The $3d$ plaquette Ising Hamiltonian, with the Ising spins $\sigma=\pm 1$ sited
at the vertices of a $3d$ cubic lattice,
\begin{equation}
  \label{e2k}
  H =  -  \frac{1}{2} \sum_{[i,j,k,l]}\sigma_{i} \sigma_{j}\sigma_{k} \sigma_{l}
\end{equation}
sits at the  $\kappa=0$ limit of a one-parameter family of $3d$ ``gonihedric''
Ising Hamiltonians \cite{savvidy,3a,Johnston2008}. In general these contain
nearest neighbour $\langle i,j \rangle$, next-to-nearest neighbour $\langle
\langle i,j \rangle \rangle$ and plaquette interactions $[i,j,k,l]$,
\begin{equation}
  H^\kappa = -2\kappa\nnsum+\frac{\kappa}{2}\nnnsum-\frac{1-\kappa}{2}\plaqsum\;,
  \label{eq:ham:goni}
\end{equation} 
which have been fine tuned to eliminate the bare surface tension. While not
gauge theories, the Hamiltonians are still highly symmetric.  In the general
case when $\kappa \ne 0$, parallel, non-intersecting planes of spins may be
flipped in the ground state at zero energy cost, which gives a $3 \times
2^{2L}$ ground-state degeneracy on an $L \times L \times L$ cubic lattice. A
low-temperature expansion reveals that this is broken at finite
temperature \cite{pietig_wegner}. However, when $\kappa=0$, planes of spins
(including intersecting ones) may be flipped throughout the low-temperature
phase, giving a macroscopic low-temperature phase degeneracy of $2^{3L}$.  This
in turn leads to non-standard corrections to finite-size scaling at the
first-order transition in the model \cite{goni_prl,goni_muca,goni_athens}. The
same degeneracy precludes the use of a standard magnetic order parameter such
as $m = \sum_i \sigma_i/L^3$ in the plaquette model since it would be zero
throughout the low-temperature phase.

In this paper we identify a class of suitable order parameters for the
$3d$ plaquette gonihedric Ising model and using high-precision multicanonical
simulations we investigate the scaling properties of the order parameters and
their associated susceptibilities.

\section{Model and Observables}

If we allow for anisotropic couplings the Hamiltonian of the $3d$ Ising model
with purely plaquette interactions may be written as 
\bea
H_{\rm aniso} &=& - J_x  \sum_{x=1}^{L} \sum\limits_{y=1}^{L}\sum\limits_{z=1}^{L} \sigma_{x,y,z} \sigma_{x,y+1,z}\sigma_{x,y+1,z+1} \sigma_{x,y,z+1} \nonumber \\
&{}&  - J_y \sum_{x=1}^{L} \sum\limits_{y=1}^{L}\sum\limits_{z=1}^{L}  \sigma_{x,y,z} \sigma_{x+1,y,z}\sigma_{x+1,y,z+1} \sigma_{x,y,z+1}  \\
&{}& - J_z \sum_{x=1}^{L} \sum\limits_{y=1}^{L}\sum\limits_{z=1}^{L}  \sigma_{x,y,z} \sigma_{x+1,y,z}\sigma_{x+1,y+1,z} \sigma_{x,y+1,z} \nonumber 
\eea
where we have indicated each site and directional sum explicitly, assuming we
are on a cubic $L \times L \times L$ lattice with periodic boundary conditions,
i.e. $\sigma_{L+1, y, z} = \sigma_{1,y,z}$, $\sigma_{x, L+1, z} = \sigma_{x,1,z}$, $\sigma_{x,y,L+1} = \sigma_{x,y,1}$,
for convenience in the sequel. 

A hint about the nature of the magnetic ordering in the model comes from
considering the strongly anisotropic limit where $J_z=0$.  In this case the
horizontal plaquettes have zero coupling, which Hashizume and Suzuki gave the
apt name of  the  ``fuki-nuke'' (``no-ceiling'' in Japanese)
model \cite{suzuki1,suzuki_old}. The low-temperature order in such an
anisotropic $3d$ plaquette Hamiltonian at $J_z=0$,
\bea
H_{\rm fuki\mbox{-}nuke} &=& - J_x  \sum_{x=1}^{L} \sum\limits_{y=1}^{L}\sum\limits_{z=1}^{L} \sigma_{x,y,z} \sigma_{x,y+1,z}\sigma_{x,y+1,z+1} \sigma_{x,y,z+1} \nonumber \\
&{}&  - J_y \sum_{x=1}^{L} \sum\limits_{y=1}^{L}\sum\limits_{z=1}^{L}  \sigma_{x,y,z} \sigma_{x+1,y,z}\sigma_{x+1,y,z+1} \sigma_{x,y,z+1}\;,
\eea
may be discerned by rewriting it as a stack of $2d$ nearest-neighbour Ising
models.
This can be carried out by defining bond spin variables $\tau_{x,y,z}
= \sigma_{x,y,z} \sigma_{x,y,z+1}$  at each vertical lattice bond.
The $\tau$ and $\sigma$ spins are thus related by
\be
\sigma_{x,y,z} =\sigma_{x,y,1}\times \tau_{x,y,1} \, \tau_{x,y,2} \, \tau_{x,y,3} \cdots \tau_{x,y,z-1}\;,
\ee
where to maintain a one-to-one correspondence between the $\sigma$ and $\tau$
spin configurations on a periodic lattice the $L^2$ constraints of the
$\tau$-spins, $\prod_{z=1}^L\tau_{x,y,z} = 1$, must be satisfied and respected
by an additional factor of $2^{L\times L}$ in the
partition function \cite{fukinuke}. 
%This can be carried out by defining 
%bond spin variables $\tau_{x,y,z} = \sigma_{x,y,z} \sigma_{x,y,z+1}$ at each
%vertical lattice bond which satisfy $\prod_{z=1}^L\tau_{x,y,z} = 1$, imposing 
%a number of $L^2$ constraints in total whose contribution should vanish in the 
%thermodynamic limit.\cite{castelnovo}
%The $\tau$ and $\sigma$ spins are thus related by
%%
%\be 
%\sigma_{x,y,z} =\sigma_{x,y,1}\times \tau_{x,y,1} \, \tau_{x,y,2} \, \tau_{x,y,3} \cdots \tau_{x,y,z-1}\;,
%\ee
%%
%and the partition function acquires an additional factor of $2^{L\times L}$
%to keep a \mbox{one-to-one} correspondence between the $\tau$ and $\sigma$-spins.
The resulting Hamiltonian when $J_x = J_y = 1$ is then
\be
H = - \sum\limits_{x=1}^{L}\sum\limits_{y=1}^{L}\sum\limits_{z=1}^{L} \left( \tau_{x,y,z}  \tau_{x+1,y,z} + \tau_{x,y,z} \tau_{x,y+1,z} \right)
\ee
which can be seen to be that of a stack of decoupled $2d$ Ising layers with
nearest-neighbour in-layer interactions in the horizontal planes apart from
the constraints that are expected to vanish in the thermodynamic
limit \cite{castelnovo}. 

Since each $2d$ Ising layer will magnetize independently at the $2d$ Ising
model transition temperature a suitable order parameter in a single layer is
the standard Ising magnetization, which in terms of the original $\sigma$ spins
is
\be 
m_{2d, z} = \left<  \frac{1}{L^2} \sum_{x=1}^{L} \sum\limits_{y=1}^{L} \sigma_{x,y,z} \sigma_{x,y,z+1}   \right> \; .
\ee
%%
%\be 
%m_{2d, z} = \left<  \frac{1}{L^2} \sum_{x=1}^{L} \sum\limits_{y=1}^{L} \sigma_{x,y,z} \sigma_{x,y,z+1}   \right> \; .
%\ee
%%
The suggestion of Hashizume and Suzuki \cite{suzuki1} and
Johnston \cite{goni_order} was that similar order parameters could still be
viable for the {\it isotropic} plaquette action.  To avoid accidental
cancellations between different planes we could use either absolute values for
each plane or square the values. This gives a candidate $3d$ order parameter in
the isotropic case as either
\begin{eqnarray}
  m_{\rm abs}^x = 
    \left< \frac{1}{L^3}\sum_{x=1}^{L} \left|\sum\limits_{y=1}^{L}
    \sum\limits_{z=1}^{L} \sigma_{x,y,z}\sigma_{x+1,y,z}\right| \right> \; ,
% \nonumber
\end{eqnarray}
or
\begin{eqnarray}
  m_{\rm sq}^x = 
    \left< \frac{1}{L^5}\sum_{x=1}^{L} \left(\sum\limits_{y=1}^{L}
    \sum\limits_{z=1}^{L} \sigma_{x,y,z}\sigma_{x+1,y,z}\right)^2 \right> \; ,
% \nonumber
\end{eqnarray}
where we again assume periodic boundary conditions and with obvious similar
definitions for the other directions, $m_{\rm abs, sq}^y$ and
$m_{\rm abs, sq}^z$. For the isotropic model we would expect $m_{\rm abs}^x
= m_{\rm abs}^y = m_{\rm abs}^z$ and similarly for the squared quantities,
which can form a useful consistency test in simulations.

Metropolis simulations indicated that $m_{\rm abs}^{x,y,z}$ and
$m_{\rm sq}^{x,y,z}$ as defined above might indeed be suitable order
parameters \cite{goni_order} for the isotropic plaquette model, but these were
subject to the usual problems of simulating a strong first-order phase
transition with such techniques.  Here we discuss multicanonical Monte Carlo
simulations \cite{muca,muca-wj}, combined with reweighting 
techniques \cite{reweighting},
which allow us  to carry out much higher precision measurements of $m_{\rm
abs}^{x,y,z}$ and $m_{\rm sq}^{x,y,z}$ and confirm the suitability of the
proposed order parameters.  We also investigate their scaling properties near
the first-order transition point. 

\section{Numerical Investigation}

Details of the simulation techniques may be found in Ref.~\cite{goni_muca}. A
two-step process is used, where estimates of an unknown weight function
$W(E)$ for configurations $\left\{\sigma\right\}$ with system energy
$E=H(\left\{\sigma\right\})$ are iteratively improved. This
replaces the Boltzmann weights ${e}^{-\beta E}$ that give the acceptance
rate in traditional Metropolis Monte Carlo. In the first step the weights are
adjusted so that the transition probabilities  between configurations with
different energies become constant, giving a flat energy
histogram \cite{mucaweights} as shown in Fig.~\ref{fig:pe_muca_log} below.
\begin{figure}[htb]
  \begin{center} 
    \includegraphics[width=0.55\textwidth]{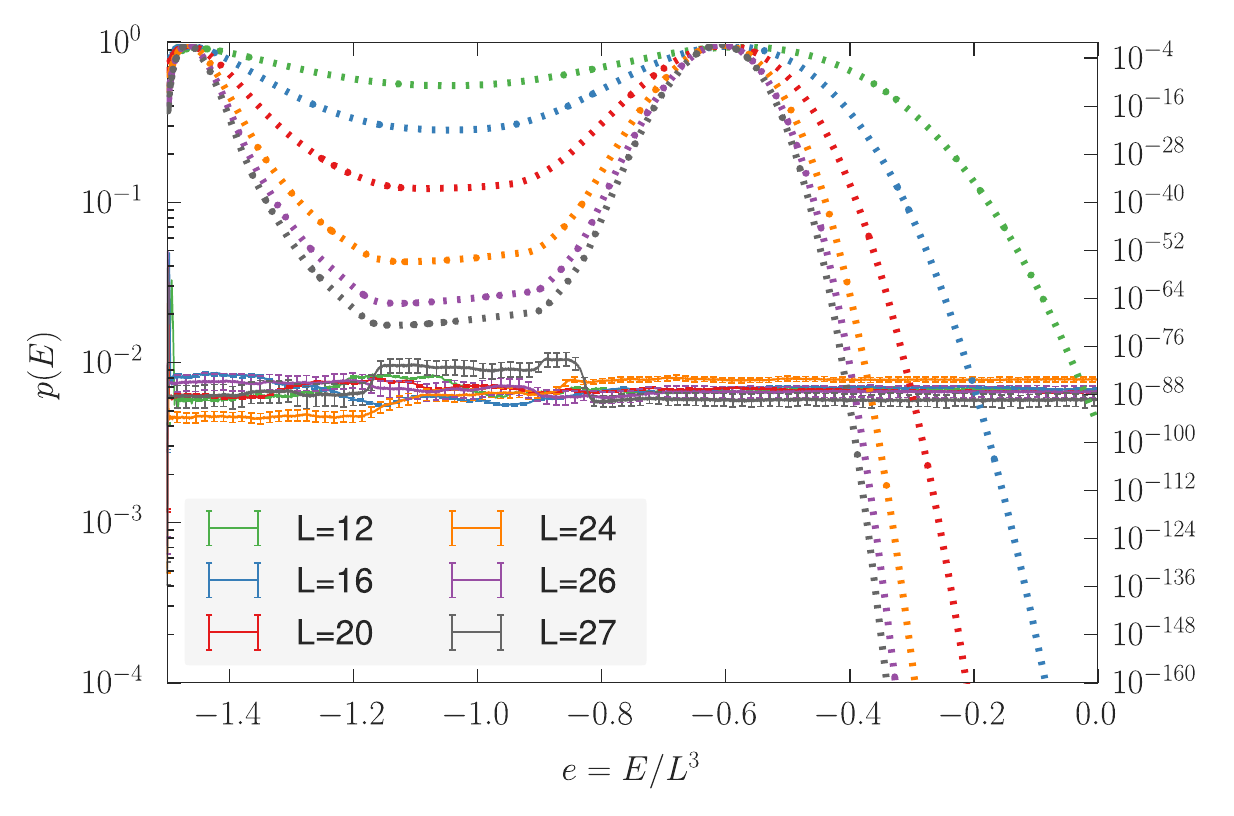} 
    \caption{The approximately flat energy histogram $p(E)$ for various lattice
      sizes produced in the multicanonical simulations on the left ordinate, and
      the reweighted canonical probability density at the temperature of equal
      peak-height on the right ordinate with dotted lines.}
    \label{fig:pe_muca_log} 
  \end{center} 
\end{figure}
The second step consists of a production run using the  fixed weights produced
iteratively in step one. This yields a time series of the energy, magnetization
and the two different fuki-nuke observables $m_{\rm abs}$ and $m_{\rm sq}$ in
the three possible different spatial orientations.  With sufficient statistics
such a time series together with the weights provides an 8-dimensional density
of states $\Omega(E, m, m_{\rm abs}^x, \dots)$ by  counting  occurrences of $E,
m, m_{\rm abs}^x, \dots$ and weighting them with the inverse $W^{-1}(E)$ of the
weights fixed prior to the production run. Practically, estimators  of the
microcanonical expectation values of observables  are used.  When measuring the
order parameters, measurements are carried out every $V=L^3$ sweeps, because
the lattice must be traversed once to measure the order parameters in all
spatial orientations which has a considerable impact on simulation times.
Skipping intermediate sweeps gives less statistics, but the resulting
measurements are  less correlated in the final time series.  

\begin{figure}[b]
	\begin{center} 
		\includegraphics[width=0.45\textwidth]{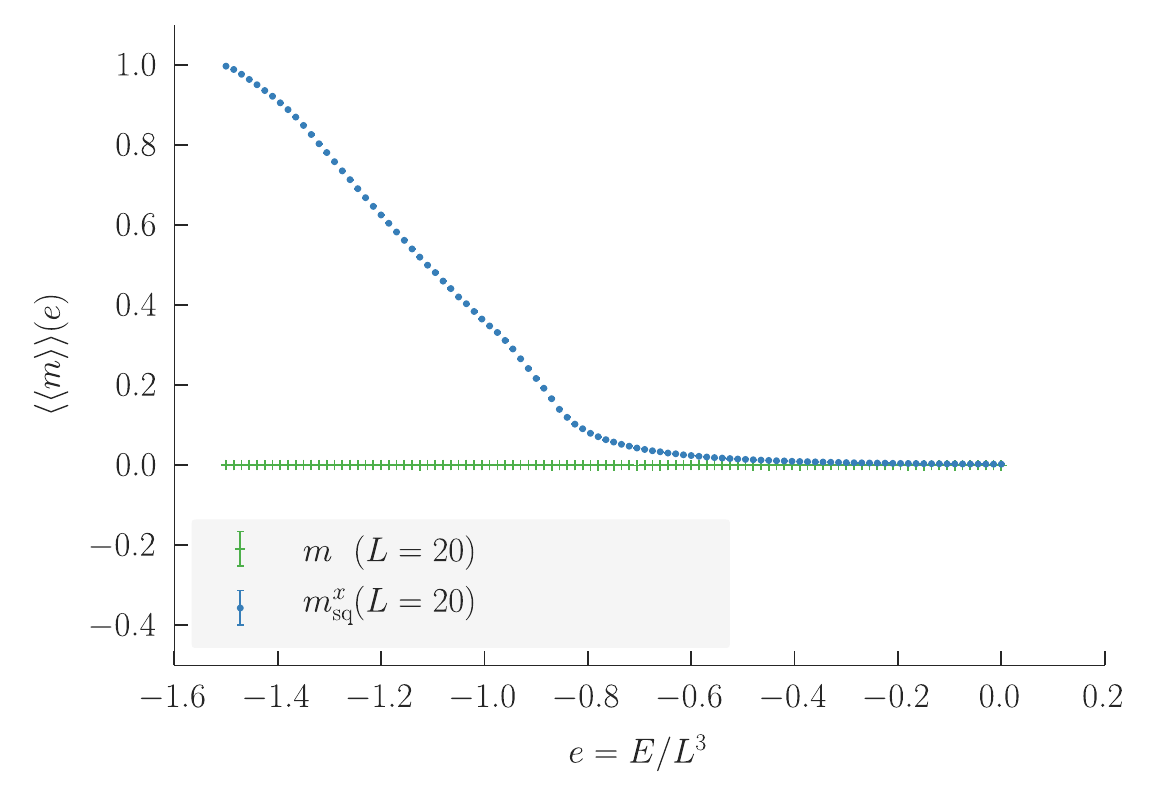} 
		\includegraphics[width=0.45\textwidth]{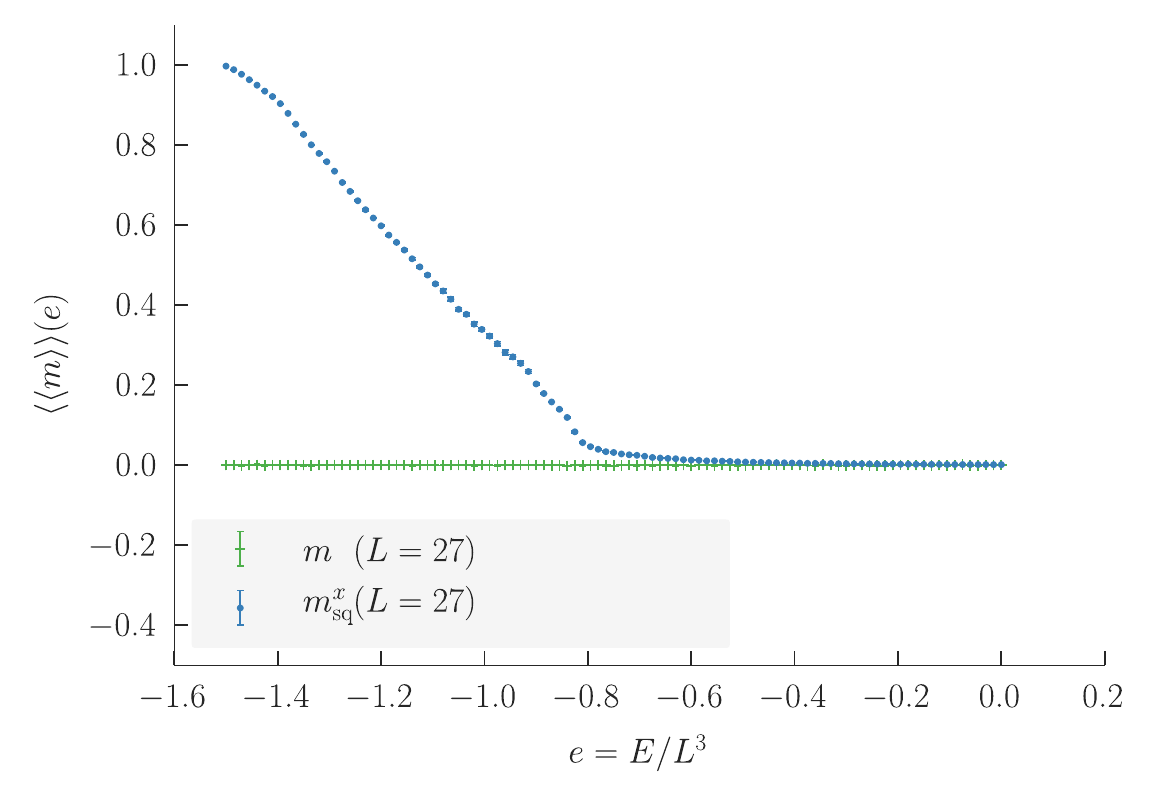} 
		\caption{Microcanonical estimators for the magnetization $m$ and fuki-nuke
			parameter $m_{\rm sq}^x$ for lattices with size $L=20$ and $L=27$, where
			we used $100$ bins for the energy normalized by system volume $e=E/L^3$ for
			this representation. Errors are obtained from Jackknife error analysis with
			$20$ blocks.}
		\label{fig:mE} 
	\end{center} 
\end{figure}

\begin{figure}[hbt] 
  \begin{center} 
    \includegraphics[width=0.45\textwidth]{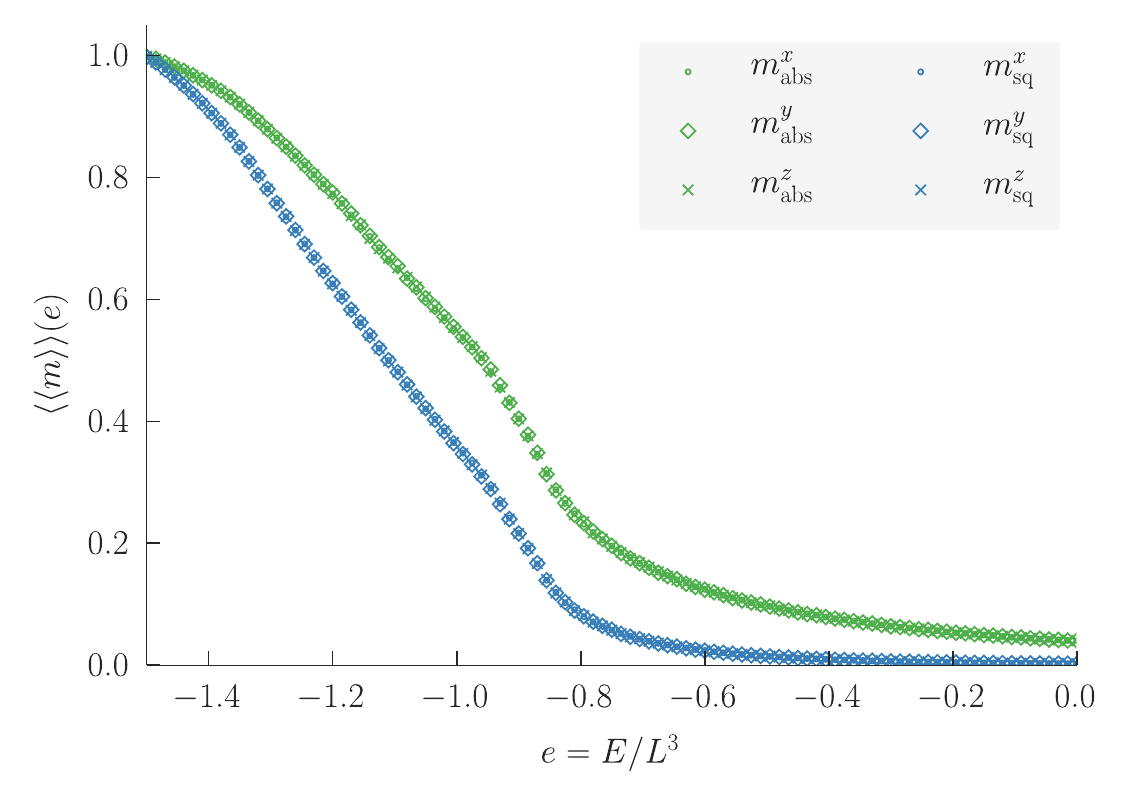}
    \includegraphics[width=0.45\textwidth]{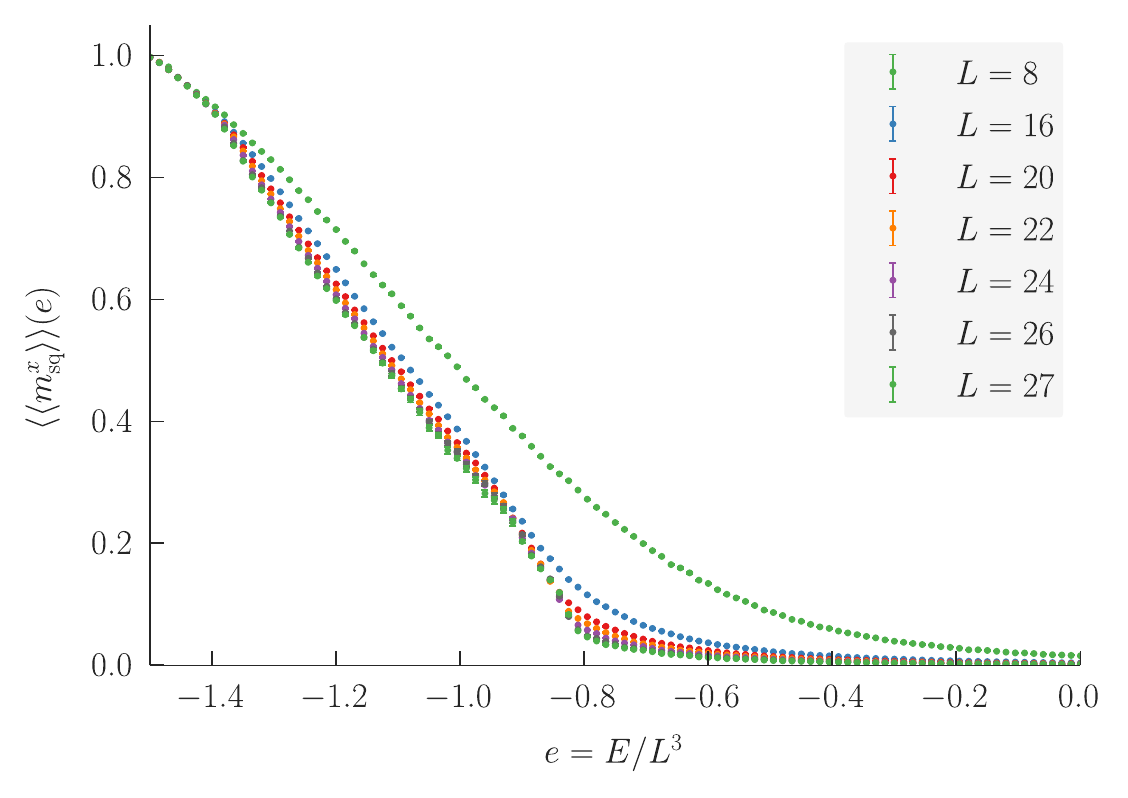} 
    \caption{Left: Microcanonical estimators for the different orientations of
      the fuki-nuke parameters $m^{x,y,z}_{\rm abs, sq}$ for a lattice with linear
      size $L=20$, which fall onto two curves. The statistical errors are smaller than the
      data symbols and have been omitted for clarity. Right: Microcanonical
      estimators for the fuki-nuke parameter $m^{x}_{\rm sq}$ for several
      lattice sizes.}
  \label{fig:microequal} 
  \end{center} 
\end{figure}

In
Fig.~\ref{fig:mE}, we show the estimators of the microcanonical expectation
values $\langle\langle \cdot \rangle\rangle$ of our magnetic observables $O$,
\begin{eqnarray}
  \langle\langle O \rangle\rangle(E) = \sum\limits_O O \, \Omega(E,O) \Big/ \sum\limits_O\Omega(E,O)\;,
  \label{eq:micro-magnetic}
\end{eqnarray}
where the quantity $\Omega(E,O)$ is the number of states with energy $E$ and
value $O$ of  either the magnetization $m$ or one of the fuki-nuke order
parameters. 
We get an estimator for $\Omega(E,O)$ by simply counting the
occurrences of the pairs $(E,O)$ in the time series and weighting them with
$W^{-1}(E)$. For clarity in the graphical representation in
\mbox{Figs.~\ref{fig:mE} and \ref{fig:microequal}} we  used a partition of
$100$ bins for the energy interval and an estimate for the statistical error of each bin
was calculated by using Jackknife error analysis \cite{jackknife} with $20$
blocks of the time series.

The fuki-nuke parameters are
capable of distinguishing ordered and disordered states, unlike the standard
magnetization, which is revealed by \mbox{Fig.~\ref{fig:mE}}.
It is also obvious from the first of Fig.~\ref{fig:microequal} that the
different orientations of the fuki-nuke parameters are equal for the isotropic
gonihedric Ising model, which we show for $L=20$. This confirms that the
sampling is consistent in the simulation.  We collect the microcanonical
estimators for $m^x_{\rm sq}$ for several lattice sizes in the second of
Fig.~\ref{fig:microequal}. 
With the stored full time series and  its weight function, we are able to
measure the microcanonical estimators for arbitrary functions of the measured
observables $f(O)$,
\begin{eqnarray}
  \langle\langle f(O) \rangle\rangle(E) = \sum\limits_O f(O) \, \Omega(E,O) \Big/ \sum\limits_O\Omega(E,O)\;,
  \label{eq:micro}
\end{eqnarray}
which provides a convenient way of calculating higher-order moments. For
canonical simulations reweighting techniques \cite{reweighting} allow system
properties to be obtained in a narrow range around the simulation temperature
whose width  and  accuracy are determined by the available statistics of
typical configurations at the temperature of interest.  Since multicanonical
simulations yield histograms with statistics covering a broad range of energies
(cf. Fig.~\ref{fig:pe_muca_log}) it is possible to reweight to a correspondingly
broad range of temperatures.  
The canonical estimator at finite inverse temperature $\beta>0$ is thus obtained as
\begin{eqnarray}
  \langle O\rangle(\beta) = \sum\limits_E \langle\langle O\rangle\rangle (E) \, {e}^{-\beta E} \Big/ \sum\limits_E {e}^{-\beta E}\;,
\end{eqnarray}
and Jackknife error analysis is again employed for an estimate of the statistical error. 

\begin{figure}[h]
	\begin{center} 
		% old plot:
		% \includegraphics[width=0.45\textwidth]{pics/mx_Ls} 
		\includegraphics[width=0.45\textwidth]{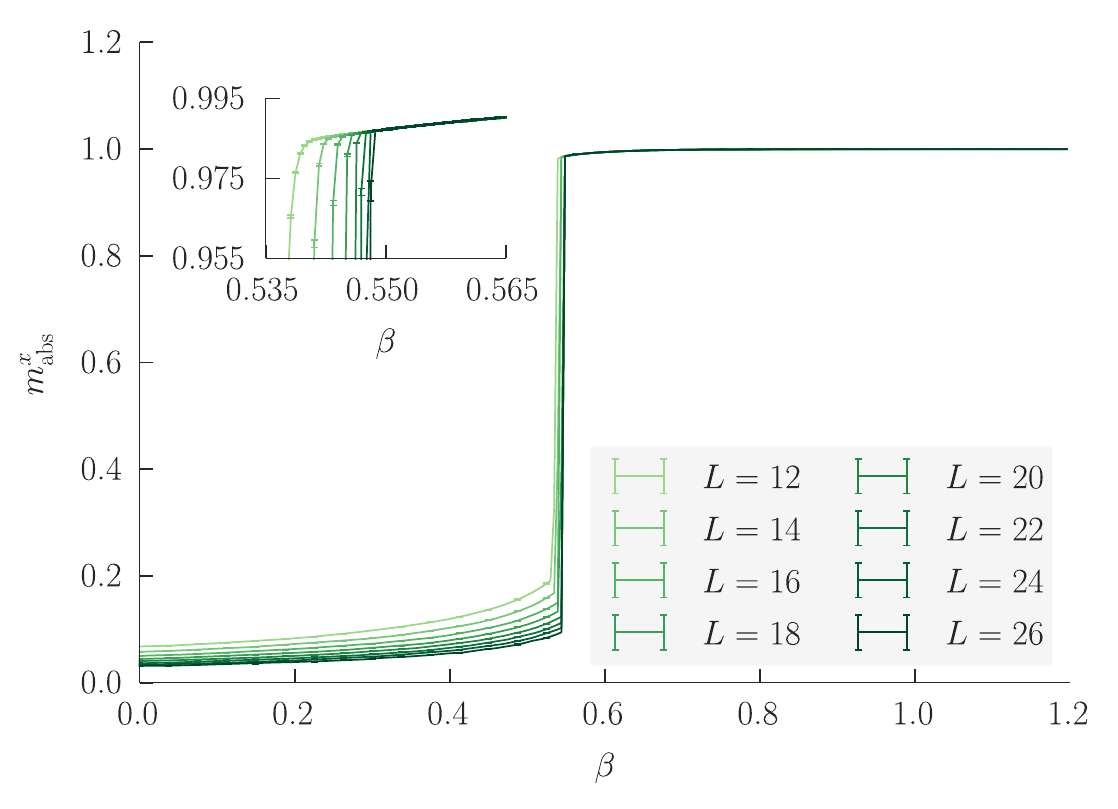} 
		\includegraphics[width=0.45\textwidth]{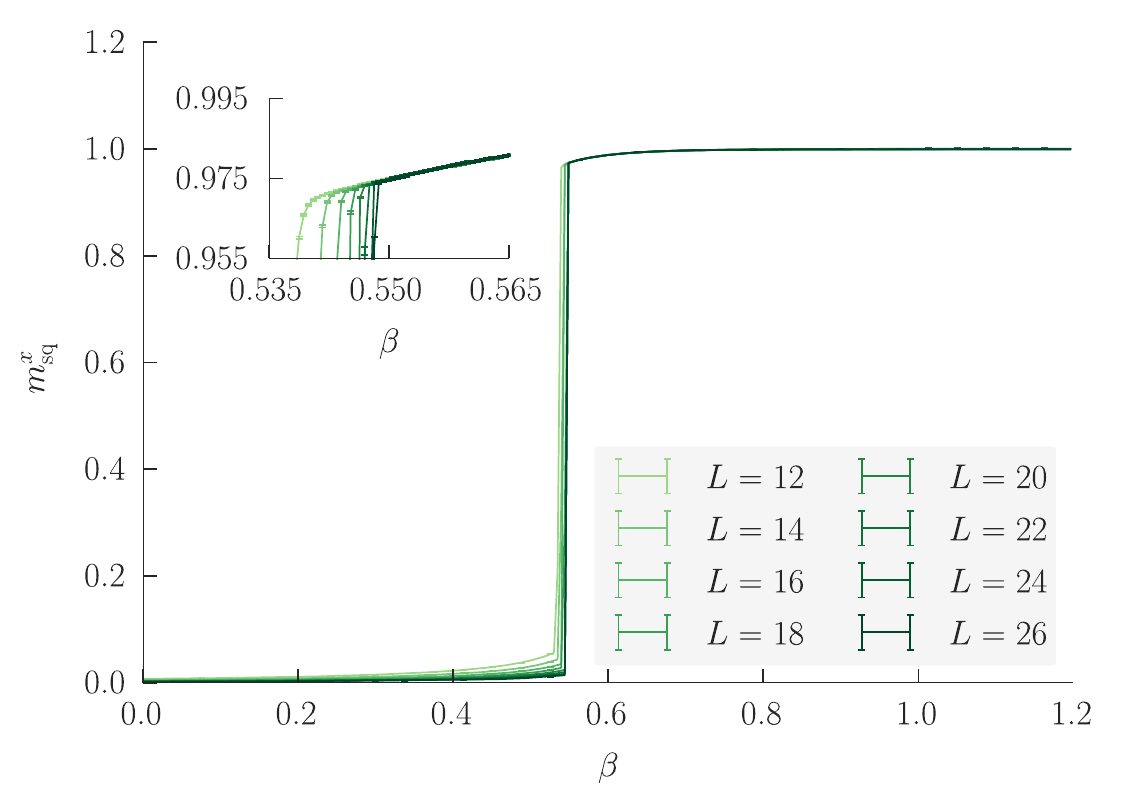}
		\caption{Canonical curves for the fuki-nuke parameters $m^{x}_{\rm abs}$
			and $m^{x}_{\rm sq}$ over a broad range of inverse temperature $\beta$
			for several lattice sizes $L$ with the insets magnifying the sharp
			jump at the ordered phase.}
		\label{fig:cmp2012} 
	\end{center} 
\end{figure}

The  behaviour of $m^{x}_{\rm abs}$ and $m^{x}_{\rm sq}$ in a Metropolis
simulation \cite{goni_order} is reproduced by the multicanonical data here as
shown in Fig.~\ref{fig:cmp2012}. 
\begin{figure}[h]
	\begin{center} 
		\includegraphics[width=0.55\textwidth]{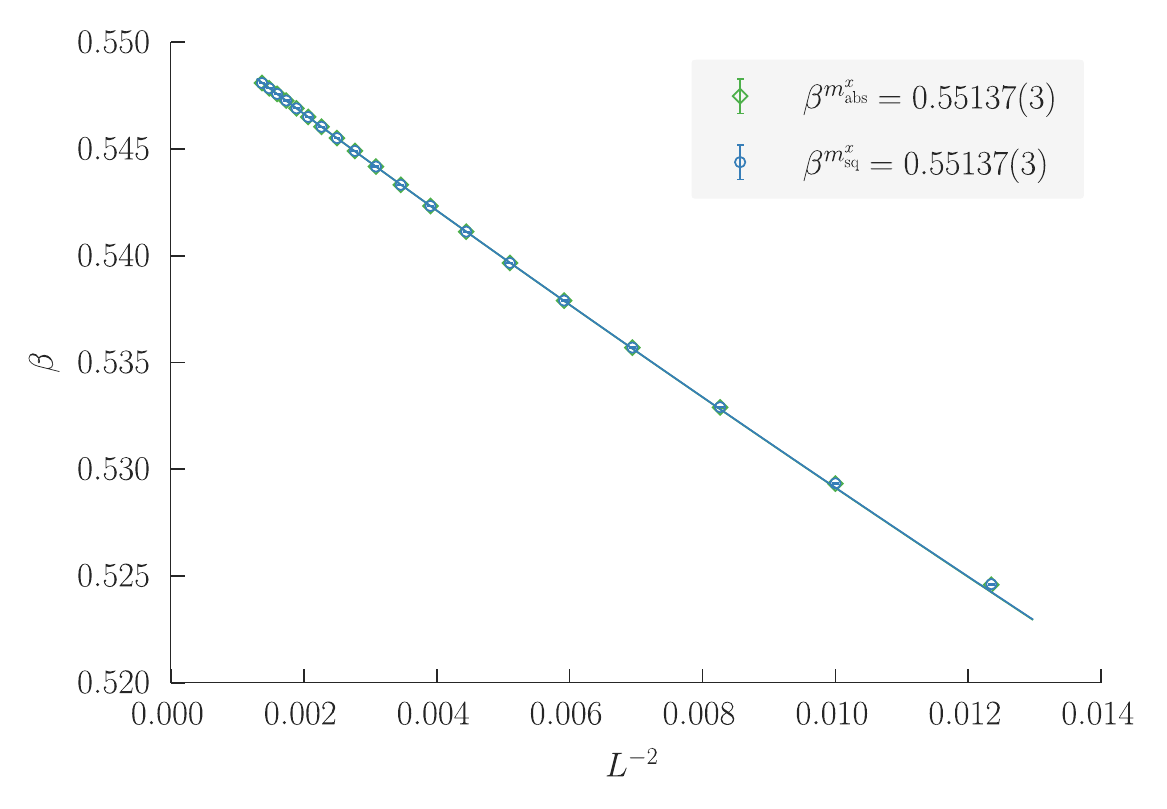} 
		\caption{Inverse temperature $\beta(L)$ as determined from the peaks in $\chi$ plotted
			against $1/L^2$ (as appropriate for the modified scaling expected with
			macroscopic degeneracy).}
		\label{fig:fukifits} 
	\end{center} 
\end{figure}
Sharp jumps {are found} near the inverse
transition temperature,  as expected for an order parameter at a first-order
phase transition. The peaks in the associated susceptibilities $\chi=\beta
L^3\,{\rm var}(m)$ provide a suitable estimate of the finite-size
pseudo-transition point. We find that the peak locations for the different
lattice sizes $L$  are  fitted best by the modified first-order scaling laws
with a leading $1/L^2$ correction appropriate for macroscopically degenerate
systems discussed in detail in Refs.~\cite{goni_prl} and
\cite{goni_muca},
\begin{eqnarray}
\label{eq:chiscaling}
  \beta^{\chi}%^{\chi_{m_{\rm abs}^{x}}}
  (L) = 0.551\,37(3) - 2.46(3)/L^2 + 2.4(3)/L^3\;,
\end{eqnarray}
where smaller lattices were systematically omitted until a fit with
quality-of-fit parameter $Q$ bigger than $0.5$ was found. The fits presented
have a goodness-of-fit parameter $Q=0.64$ and $12$ degrees of freedom left.
Fits to the other directions { $m_{\rm abs}^{y,z}$} and fits to the peak
location of the susceptibilities of $m_{\rm sq}^{x,y,z}$ give  the same
parameters within error bars and are of comparable quality.

The inverse
temperature $\beta(L)$ for the peak locations of the susceptibilities for both
$m_{\rm abs}^x$ and $m_{\rm sq}^x$ is plotted against $1/L^2$ in
Fig.~\ref{fig:fukifits} together with the best fit curve giving the quoted
values for the scaling coefficients.
The estimate of the phase transition temperature obtained here from the finite-size 
scaling of the fuki-nuke order parameter(s), $\beta^\infty =
0.551\,37(3)$, is  in good agreement  with the earlier estimate $\beta^\infty =
0.551\,334(8)$ reported in~Ref.~\cite{goni_muca} using fits to the peak location of
Binder's energy cumulant and specific heat and the value of $\beta$ where the
energy probability density has two peaks of the same height or weight. 
\begin{figure}
  \begin{center} 
    \includegraphics[width=0.45\textwidth]{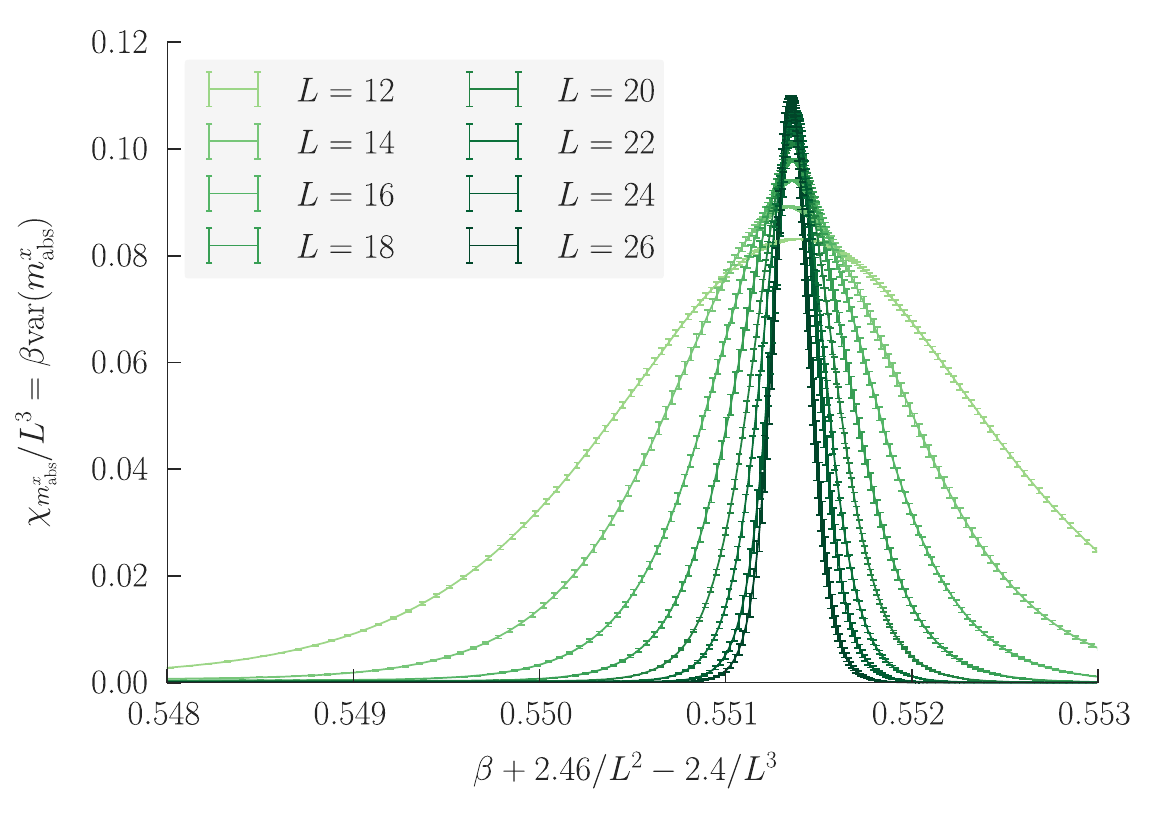} 
    \includegraphics[width=0.45\textwidth]{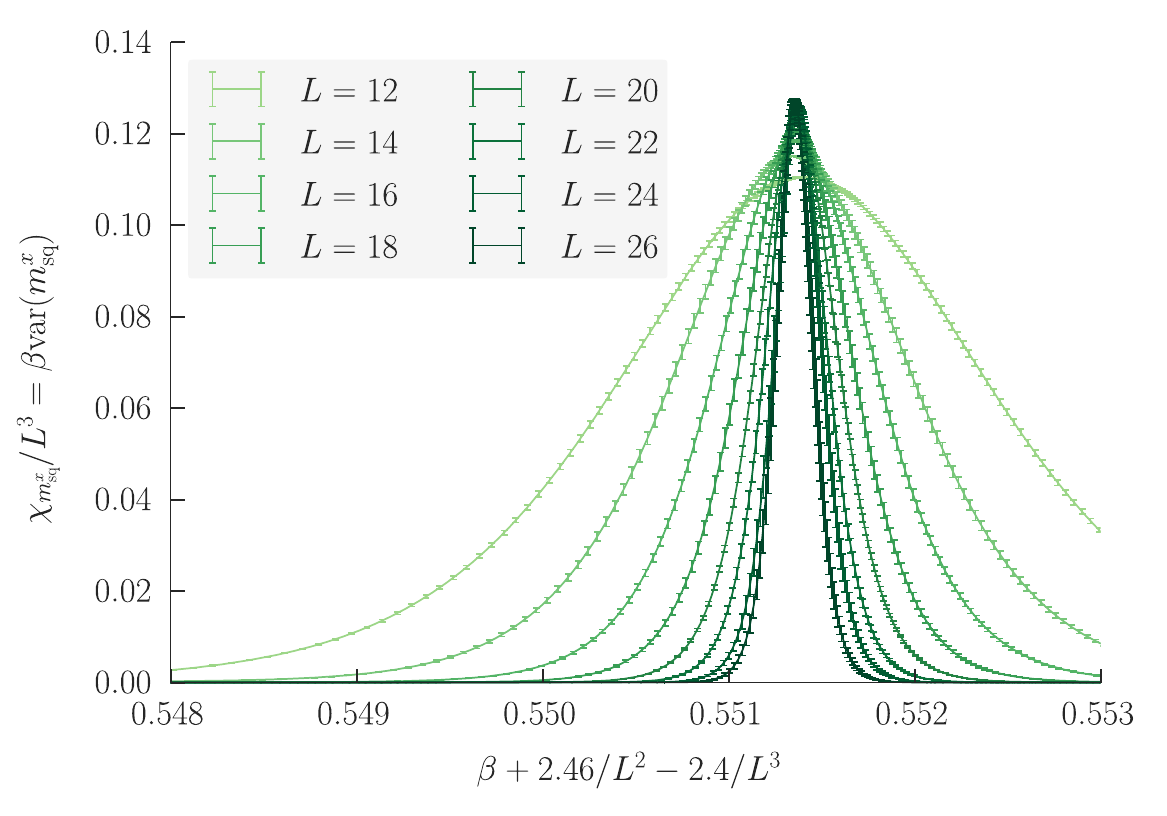}
    \caption{Canonical curves for the susceptibilities associated with the
      fuki-nuke parameters $m^{x}_{\rm abs}$ and $m^{x}_{\rm sq}$ plotted
      against the shifted inverse temperature $\beta + 2.46/L^2 -
      2.4/L^3$ for several lattice sizes $L$. The peaks fall onto the same point.}
  \label{fig:canonmx} 
  \end{center} 
\end{figure} 

\begin{figure}[b]
	\begin{center} 
		\includegraphics[width=0.75\textwidth]{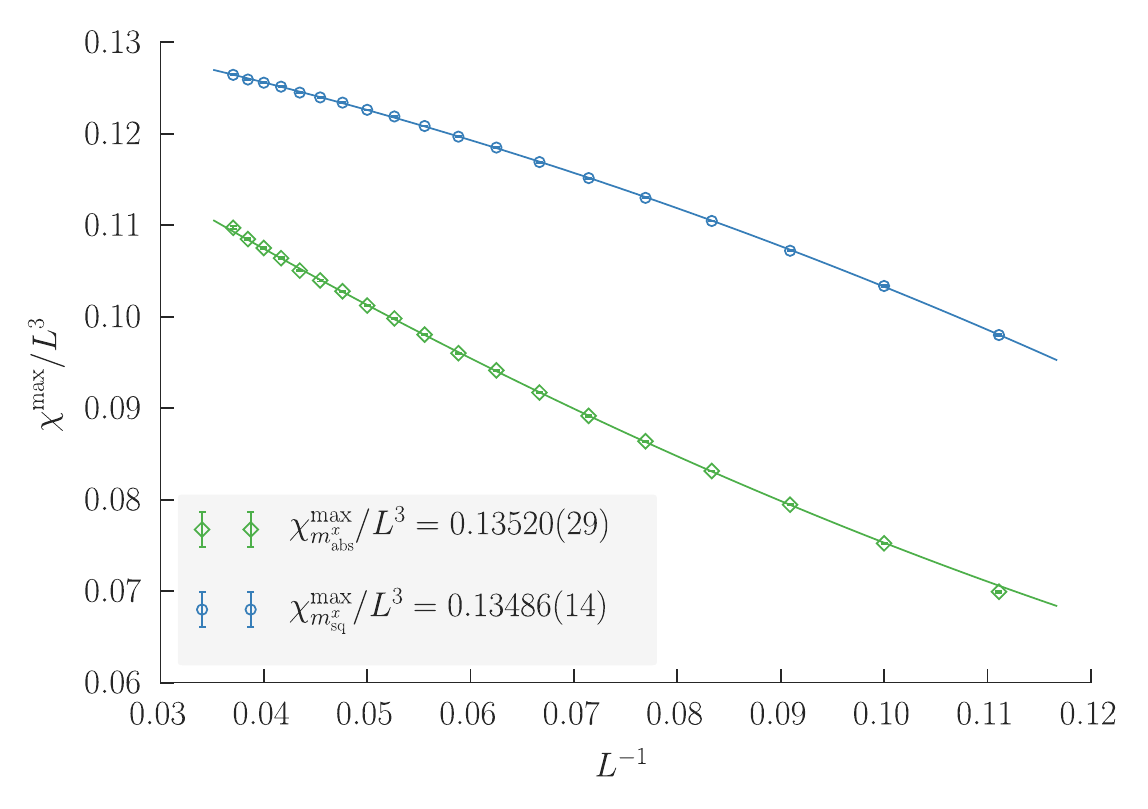} 
		\caption{Maximum value of the normalized susceptibility $\chi/L^3$ plotted
			against $1/L$ for both $\chi_{m_{\rm abs}^x}$ and $\chi_{m_{\rm sq}^x}$.}
		\label{fig:fukifitschi} 
	\end{center} 
\end{figure}

As a visual confirmation of the scaling, the normalized susceptibilities
$\chi/L^3$ for various lattice sizes are plotted against \mbox{$ \beta +
2.46/L^2 - 2.4/L^3$} (where the shifts are determined by the fitted scaling
corrections) in Fig.~\ref{fig:canonmx}. It is clear that the peak positions
fall on the same point. 
In Fig.~\ref{fig:fukifitschi} we plot the peak values of both $\chi_{m_{\rm
abs}^x}$ and $\chi_{m_{\rm sq}^x}$ divided by the system volume against $1/L$.
Empirically the plotted fits for the maximum values are 
% $$
\be
\chi_{m_{\rm abs}^x}^{\rm max}(L)/L^3 = 0.135\,20(29) -0.757(10)/L + 1.58(7)/L^2
\ee
% $$ 
where we
used all lattices with sizes greater than or equal to $L_{\rm min} = 10$ (with
$15$ degrees of freedom left and a goodness-of-fit parameter of $Q =0.07$) and
% $$
\be
\chi_{m_{\rm sq}^x}^{\rm max}(L)/L^3=0.134\,86(14) -0.174(4)/L -1.413(27)/L^2
\ee
% $$
with $L_{\rm min} = 8$ (with $17$ degrees of freedom and $Q =0.08$). The
leading corrections of the peak value of the specific heat divided by the
system volume are of order $O(1/L^2)$, so those for the susceptibility are
already much stronger. Forcing a fit to leading $1/L^2$ corrections here gives
slightly poorer fits and we cannot distinguish empirically. The constants,
however, are stable around $\chi^{\rm max}/L^3\approx0.135$, and barely
change with the various fits, showing that the proportionality factor
of $\chi^{\rm max} \propto L^3$ is independent of the leading corrections. By 
analogy with the maximum of the specific heat, \cite{goni_prl,goni_muca,goni_athens} 
we expect for the extremal values of the susceptibilities
\begin{equation}
  \chi^{\rm max}_{\rm abs, sq}/L^3 = \beta^\infty\left(\frac{\Delta \widehat{m}_{\rm abs, sq}}{2} \right)^2 + \dots\;,
\end{equation}
where the gaps $\Delta\widehat{m}_\mathrm{abs, sq} =
\widehat{m}^\mathrm{ordered}_\mathrm{abs, sq} -
\widehat{m}^\mathrm{disordered}_\mathrm{abs, sq}$ of the fuki-nuke order
parameters in the infinite system enter. This is  consistent with the
measurements since a value of $\chi^\mathrm{max}/L^3 \approx 0.135$ implies 
$\Delta\widehat{m}_\mathrm{abs, sq} \approx 0.99$, which is plausible from the insets
in Fig.~\ref{fig:cmp2012} giving an impression of the value of
$\widehat{m}^\mathrm{ordered}_\mathrm{abs, sq}$ as $L\rightarrow\infty$ and
noting that $\widehat{m}^\mathrm{disordered}_\mathrm{abs, sq} = 0$.

\section{Conclusions}

The macroscopic degeneracy of the low-temperature phase in the $3d$ plaquette
gonihedric Ising model excludes standard magnetic ordering. However,
consideration of the strongly anisotropic limit of the model suggests that a
planar, fuki-nuke order may still be present.  Multicanonical simulations of
the model strongly support this suggestion, with the various fuki-nuke
magnetizations all showing order parameter like behaviour. If the effects of
the macroscopic low-temperature phase degeneracy on the corrections to scaling
detailed in Refs.~\cite{goni_prl} and \cite{goni_muca} are taken into
account, the estimates for the transition point obtained from these fuki-nuke
magnetizations are fully consistent with estimates obtained from energetic
quantities. There are, however, stronger finite-size corrections to the peak
value of the susceptibilities for the fuki-nuke order parameters.  

\section*{Acknowledgements}
This work was supported by the Deutsche Forschungsgemeinschaft (DFG)
through the Collaborative Research Centre SFB/TRR 102 (project B04) and by
the Deutsch-Franz\"osische Hochschule (DFH-UFA) under Grant No.\ CDFA-02-07.

\end{document}